\documentclass{epl}

\usepackage{epsfig}

\title{Dynamics of sliding drops on superhydrophobic surfaces}

\author{A.~Dupuis\inst{1,2} \and J.M.~Yeomans\inst{1}}
\institute{
  \inst{1} The Rudolf Peierls Centre for Theoretical Physics, University of Oxford, 1 Keble Road, Oxford OX1 3NP, UK. \\
  \inst{2} Institute of Computational Science, ETH Zurich, 8092 Zurich, Switzerland.
}
 
\shortauthor{Dupuis and Yeomans}

\begin{document}

\maketitle

\begin{abstract}
We use a free energy lattice Boltzmann approach to investigate
numerically the dynamics of drops moving across superhydrophobic
surfaces. The surfaces comprise a regular array of posts small
compared to the drop size. For drops suspended on the posts the
velocity increases as the number of posts decreases. We show that this
is  because the velocity is  primarily determined by the contact angle
which, in turn, depends on the area covered by posts.  Collapsed
drops, which fill the interstices  between the posts, behave in a very
different way. The posts now impede the drop behaviour and the
velocity falls as their density increases.
\end{abstract}

\newcommand{\pos}{{\mathbf{r}}}
\newcommand{\eq}[1]{equation (\ref{#1})}
\newcommand{\Eq}[1]{Equation (\ref{#1})}
\newcommand{\fig}[1]{fig.~\ref{#1}}
\newcommand{\Fig}[1]{Fig.~\ref{#1}}

% ------------------------------------------------------------
\section{Introduction}

The aim of this letter is to explore numerically how micron-scale 
drops move on
superhydrophobic surfaces. If surfaces with contact angles greater
than $90^o$ are patterned with posts small compared to the drop
dimensions the equilibrium contact angle is increased and can reach
values close to $180^o$. Such superhydrophobic surfaces are found in
nature: for example the leaves of several plants, such as the lotus
are covered in tiny bumps which may
have evolved to aid the run-off of rain
water. Recent microfabrication techniques have
allowed superhydrophobic patterning to be
mimicked and carefully controlled
experiments on the behaviour of drops on superhydrophobic substrates
are increasingly becoming feasible\cite{quere:05}.

Drops on superhydrophobic surfaces can be in two states. 
Suspended drops lie on top of the
posts with air pockets beneath them whereas collapsed drops fill the
interstices between the posts. A suspended drop has a higher contact
angle than the equivalent collapsed drop which in turn has a higher
contact angle than a drop on a flat surface made of the same
material. Several authors have shown that the equilibrium properties
of the drops follow from thermodynamic arguments based on free energy
minimisation~\cite{ishino:04,patankar:04,marmur:03,swain:98}. Both %extrand:02
the suspended and collapsed states can provide the global minimum of
the free energy, with the phase boundary between them depending on the
surface tension, the contact angle on the flat surface and the post
geometries~\cite{cottin-bizonne:03,journet:05}. The suspended drop may
often exist as a metastable state as it has to cross a free energy
barrier to fill the grooves~\cite{patankar:04,dupuis:04e}.

There is, however, no similar understanding of the way drops move
across superhydrophobic surfaces. Several
authors~\cite{krasovitski:05,oner:00,mchale:04} have considered
contact angle hysteresis in the context
of whether a drop is pinned or moves as a surface is tilted. The
consensus of both theoretical and experimental work is that a
suspended drop on a superhydrophobic surface starts to move much more
easily than the equivalent drop on a smooth surface. 

Only a few results have been published about the steady-state
velocities of moving drops. Gogte {\it et~al.}~\cite{gogte:05} have
shown that the drag on a hydrofoil decreases by $\sim 10 \%$ if it is
covered by a superhydrophobic coating.  Ou {\it et~al.}~\cite{ou:04}
have shown experimentally that a drag reduction $\sim 40 \%$ can be
obtained for drops on surfaces with micron-scale posts. Cottin-Bizonne
{\it et~al.}~\cite{cottin-bizonne:03} have used molecular dynamics
simulations to explore the behaviour of a fluid moving across a
superhydrophobic surface comprising nanometre posts. They find that
the slip is enhanced by a factor $\sim 2.5$ when the fluid does not
fill the space between the posts and reduced by a factor $\sim 10$
when it does, as compared to a flat surface of the same (hydrophobic)
material.
 
In this letter we explore movement across a superhydrophobic surface
by using a lattice Boltzmann approach to solve the equations of motion
of a drop pushed gently by a constant force. This approach has been
shown to agree well with experiments for spreading and moving drops on
chemically striped surfaces~\cite{dupuis:03c,kusumaatmaja:05}. Here we
consider both suspended and collapsed drops on superhydrophobic
surfaces and investigate how the drop velocity depends on the post
spacing.  We find that the velocity of suspended drops is essentially
determined by the effective contact angle and increases as the number
of posts decreases. Collapsed drops, however, behave in a very
different way. The posts now impede the drop behaviour and for large
post densities the flow pushes the drop back to a suspended position
on top of the posts.

% ------------------------------------------------------------
\section{The model}

We consider a liquid-gas system of density $n(\pos)$ and volume
%$V$. The surface of the substrate is denoted by $S$. The equilibrium
$V$. The surface is denoted by $S$. The equilibrium properties of the
drop are described by the Ginzburg-Landau free energy
\begin{equation}
\Psi = \int_V \left( \psi_b(n) + \frac{\kappa}{2} \left( \partial_\alpha n
  \right)^2 \right) dV + \int_S \psi_c(n) \; dS
\label{eq:freeE}
\end{equation}
where Einstein notation is understood for the Cartesian label
$\alpha$. $\psi_b(n)$ is the free energy in the bulk. For convenience
we choose a double well form.
%~\cite{briant:04}. 

The derivative term in \eq{eq:freeE} models the free energy associated
with density gradients at an interface. $\kappa$ is related to the
surface tension. $\psi_c(n_s)=-\phi_1 n_s$, where $n_s$ denotes the
density at the surface, is the Cahn surface free energy~\cite{cahn:77}
which controls the wetting properties of the fluid. In particular
$\phi_1$ can be used to tune the contact angle.
 
The dynamics of the drop is described by the Navier-Stokes equations
for a non-ideal gas
\begin{eqnarray}
\partial_t (n u_\alpha) + \partial_\beta (nu_\alpha u_\beta) & = & 
  - \partial_\beta P_{\alpha\beta} 
  + \nu \partial_\beta \big[ n (\partial_\beta u_\alpha + 
                              \partial_\alpha u_\beta + 
                              \delta_{\alpha\beta} \partial_\gamma u_\gamma) 
                       \big] + nF_\alpha,
\label{eq:ns}  \\
%\end{equation*}
%\begin{equation}
\partial_t n + \partial_\alpha(n u_\alpha)  & = &  0
\label{eq:cont} 
\end{eqnarray}
where $\mathbf{u}(\pos)$ is the fluid velocity, $\nu$ the kinematic
viscosity and $n \mathbf{F}$ a body force per unit volume. In what follows, we
will refer the second term on the right hand side of the first
equation as the viscous term. The pressure tensor $P_{\alpha\beta}$ is
calculated from the free energy~\cite{swift:96}.
%
%The pressure tensor $P_{\alpha\beta}$ is
%related to the free energy by~\cite{anderson:98}
%%
%\begin{equation}
%P_{\alpha\beta} = \frac{\partial \mathcal{F}}{\partial (\partial_\alpha n)}
% (\partial_\beta n) - \mathcal{F} \delta_{\alpha\beta}
%\end{equation}
%%
%where $\mathcal{F}=\psi_b - \mu_b n + \kappa(\partial_\alpha n)^2/2$
%and $\mu_b=4p_c(1-\beta\tau_w)/n_c$ is the bulk chemical potential.
%
%We use a lattice Boltzmann model to solve equations (\ref{eq:ns}) 
%and (\ref{eq:cont}). Details can be found in~\cite{dupuis:04e,swift:96}. 

We use a lattice Boltzmann algorithm that solve the equations of
motion by following the evolution of particle distribution
functions. This corresponds to a discretisation of a simplified
Boltzmann equation. Details can be found in~\cite{dupuis:04e}.

% ------------------------------------------------------------
\section{Results}

We consider a drop of radius 30 moving on a domain of size $L_x \times
L_y \times L_z = 80 \times 80 \times 80$ lattice sites. Periodic
boundary conditions are imposed along $x$ and $y$. The surface at
$z=L_z-1$ is flat and that at $z=0$ is patterned by posts in a square
array with height, spacing and width denoted by $h$, $d$ and $w$
respectively. Unless otherwise specified $h=5$. A contact angle
$\theta_0=110^o$ is set on every substrate site by choosing a suitable
value of the surface field $\phi_1$. In equilibrium the drop forms a
spherical cap with an enhanced contact angle. For example, if $d=8$
and $w=4$ the contact angle is $156^o$ for a drop resting on the posts
and $130^o$ for a drop collapsed between the posts~\cite{dupuis:04e}.

\begin{figure}
\begin{center}
\begin{tabular}{ccc}
\epsfig{file=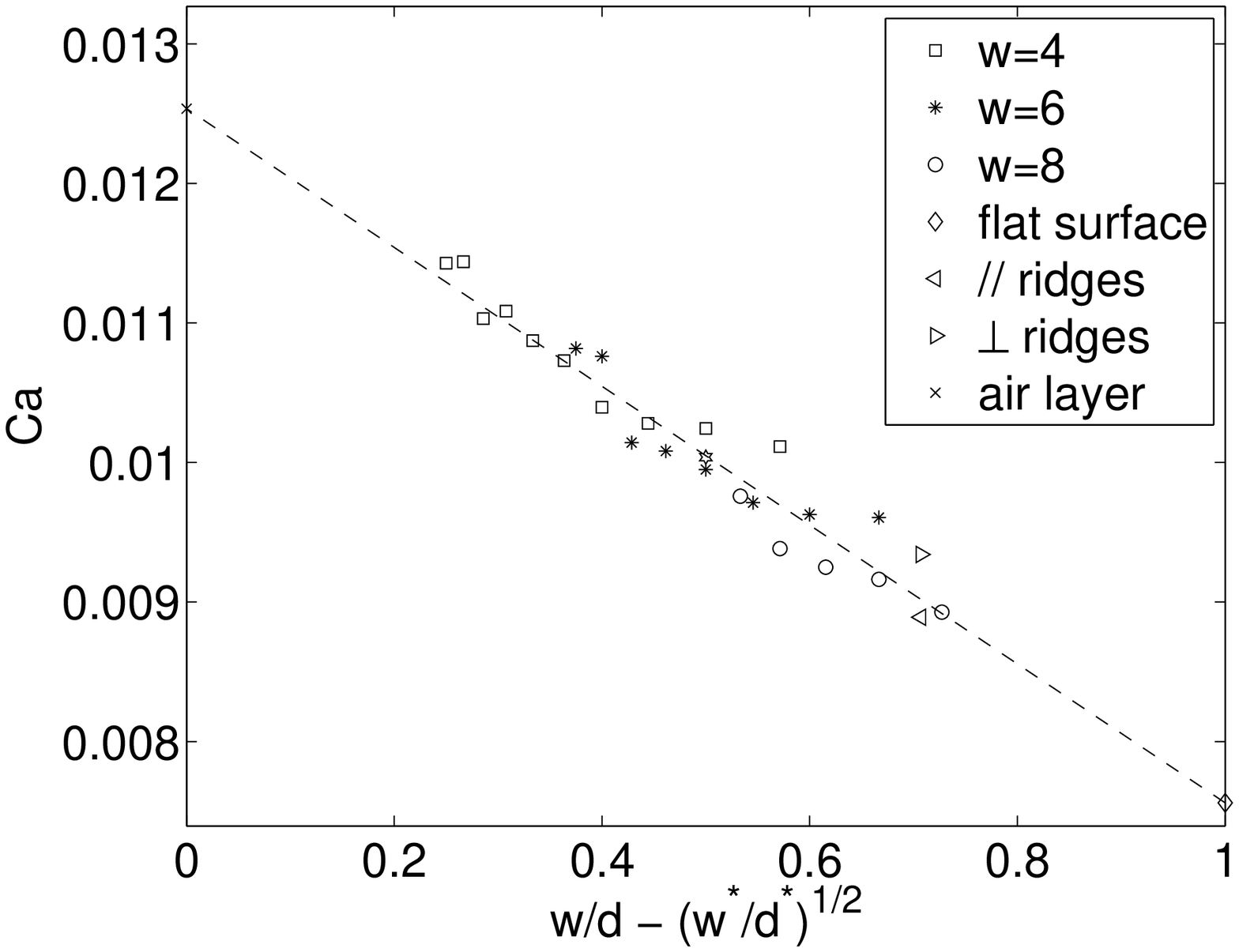,width=6.8cm} & &
\epsfig{file=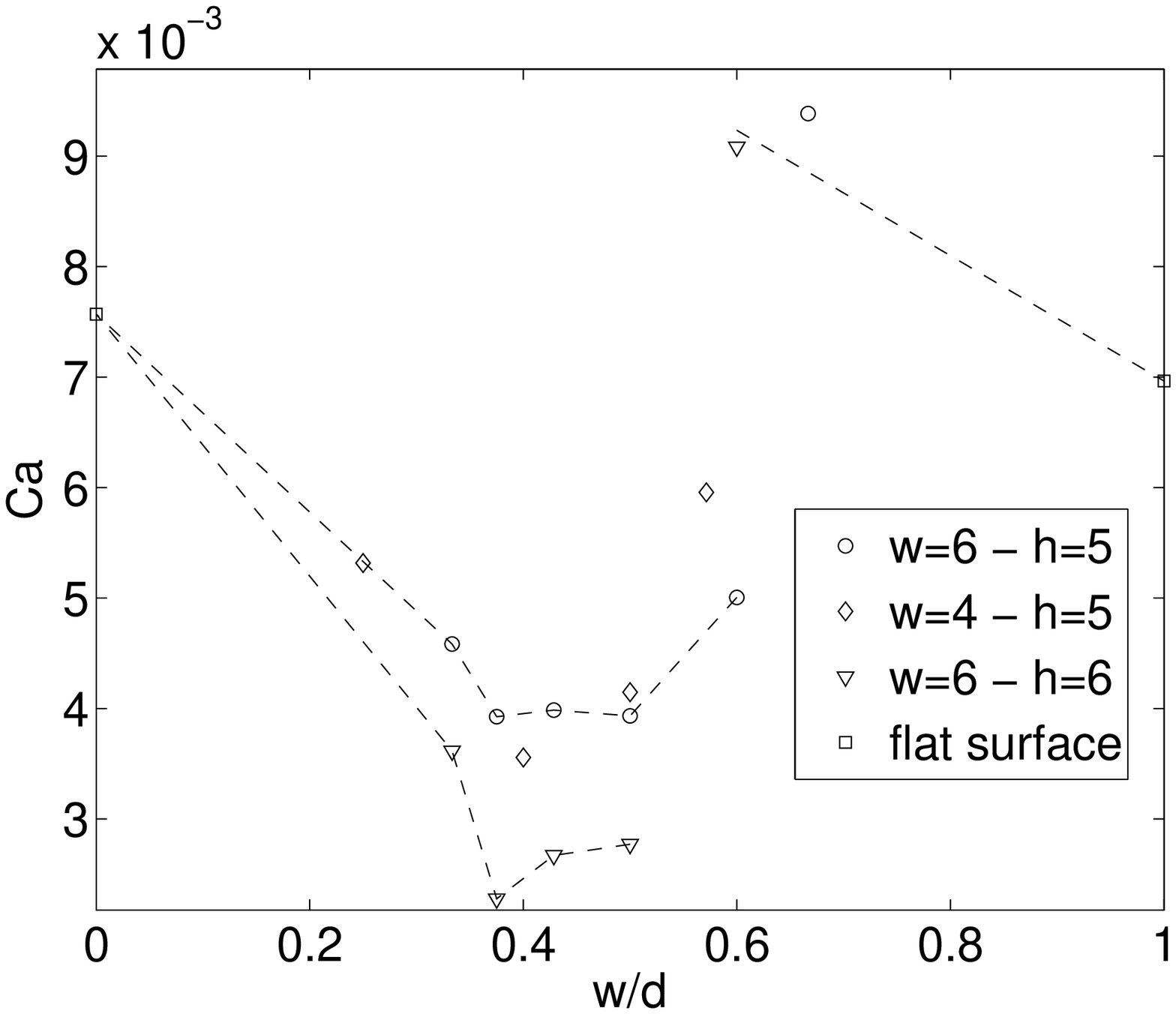,width=6.4cm} \\ (a) && (b)
\end{tabular}
\end{center}
\caption{Steady state capillary number of (a) suspended (b) collapsed
  drops.  The height, spacing and width of the posts are denoted by
  $h=5$, $d$ and $w$ respectively (in simulation units, compared to
  the drop radius $R=30$). A contact angle $\theta_0 = 110^o$ is set
  on every substrate site. We impose a Poiseuille flow with
  $Bo=1.42$. The data is plotted as a function of $(w/d)$ for surfaces
  with posts and $(w^*/d^*)^{1/2}$ when ridged surfaces are
  considered. A vertical velocity is imposed on collapsed drops for
  them to cross the suspended to collapsed free energy barrier. The
  variation in the velocity of the drop because of its position
  relative to the posts is less that $1\%$ in (a) and less than $5\%$
  in (b).}
\label{fig:velOnPosts}
\end{figure}

We impose a Poiseuille flow by considering a constant body force
$nF_x$ per unit volume parallel to the surface corresponding to a Bond
number $Bo=V^{2/3} n F_x/\gamma=1.42$ where $\gamma$ is the surface
tension and $V$ the drop volume. \Fig{fig:velOnPosts}(a) shows the
time average capillary number $Ca=\nu U n_l \gamma^{-1}$ of the
suspended drop as the width of the posts and the distance between the
posts are varied. $U$ is the velocity of the drop and $n_l$ the liquid
density. The $w/d = 0$ result corresponds to a drop moving on a
cushion of air (of height $5$), and the $w/d = 1$ value to a drop
moving on a flat surface of contact angle $110^o$. It is striking how
the data collapses onto a single curve if it is plotted as a function
of the ratio $w/d$.  We also measured the velocity of drops moving on
a ridged surface, considering both a drop moving parallel and
perpendicular to the ridges. This data also collapses onto the same
curve if plotted as a function of $(w^*/d^*)^{1/2}$, where $w^*$ is
the width of the ridges and $d^*$ the spacing between them.

\Fig{fig:velOnPosts}(b) compares the behaviour of a collapsed drop 
driven by the same force. The dependence of the velocity on the length
ratio $w/d$ is now very different.

From the results of \fig{fig:velOnPosts} we can formulate the
questions that need to be answered to understand the way in which the
drops move over the posts. Firstly why is the velocity controlled by
the ratio $w/d$ for the drops suspended on the posts? Secondly why is
the behaviour of the collapsed drop more complicated with seemingly
two different regimes as $w/d$ increases from zero to unity? To answer
these questions it is helpful to first consider in some detail the
behaviour of a drop moving on a flat surface.

%\subsection{Motion across a flat surface}

% !!! Now D=36

We therefore consider a system of size $L_x \times L_y \times L_z =
80 \times 80 \times 40$ with no-slip walls at $z=0$ and
$z=L_z-1$ and periodic boundary conditions along $x$ and $y$. A liquid
droplet is placed on the surface at $z=0$ and forms a
spherical cap of contact angle $\theta_0$.

\begin{figure}
\begin{center}
\begin{tabular}{cp{0.3cm}c}
\epsfig{file=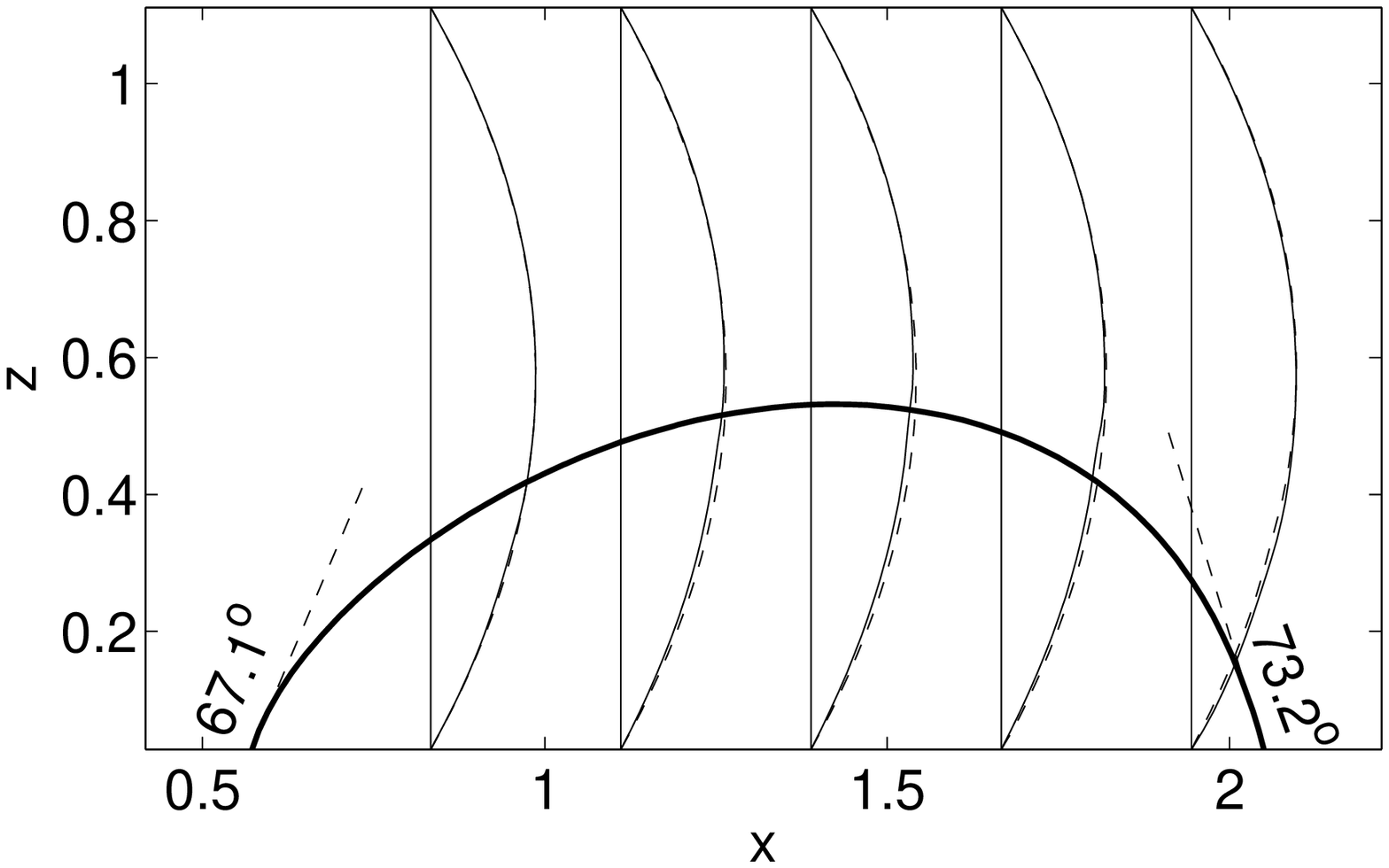,width=5.5cm} && 
  \epsfig{file=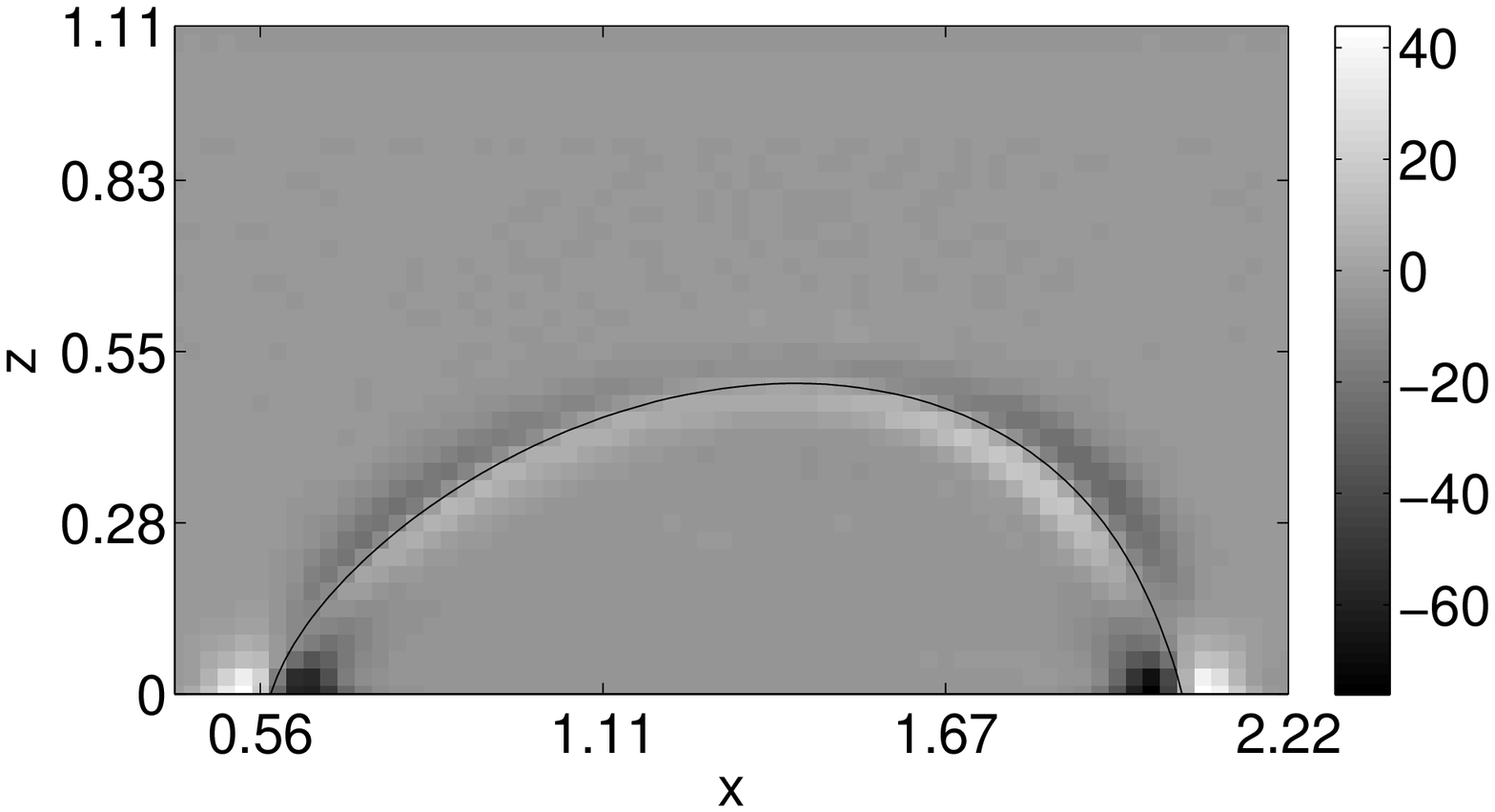,width=6.4cm} \\
(a) && (b)
\end{tabular}
\end{center}
\caption{Steady state of a sliding droplet with $\theta_0=70^o$ in a
  Poiseuille flow with $Bo=6.4$. The advancing and receding contact
  angles are $73.2^o$ and $67.1^o$. (a) Cut along $y=L_y/2$. The thick
  line is the interface where $n(\pos)=n_c$. The thin solid lines are
  velocity profiles when the drop is in place. Dashed lines are
  velocity profiles with no drop. (b) $x$-component of the viscous
  term in the Navier-Stokes equation (dimensionless units, scaled by
  $V^{(2/3)}/\gamma$) across the same cross section. Length is
  dimensionless scaled by the drop diameter $D$.}
\label{fig:vtx}
\end{figure}

We impose a Poiseuille flow by setting a body force $nF_x$
per unit volume along $x$. The
droplet becomes slightly elongated along the flow direction and
increases the front and decreases the back contact angle. For example,
the steady state of a drop
 of radius $R=18$ with $\theta_0=70^o$
pushed by a force corresponding to $Bo=6.4$ is depicted in
\fig{fig:vtx}. The front and back contact angles become $\theta_f =
73.2^o$ and  $\theta_b=67.1^o$.
This occurs because, in order that the drop can move, the contact line
has to slide along the surface, in a way that does not violate the
non-slip boundary conditions on the velocity field. This is mediated
by the advancing and receding contact angles deviating from their
equilibrium values so that the interface does not have the correct
radius of curvature. Terms in the pressure tensor act to restore the equilibrium curvature. 

%This corresponds to a
%physical process of evaporation and condensation~\cite{briant:04}.
%For a given drop velocity the advancing and receding contact angles
%adjust to values which allow the interface to attain the correct
%velocity.

Our first aim is to understand what determines the velocity of the
drop. \Fig{fig:vtx}(a) shows the flow profile with and without the
drop in place. There is very little change due to the presence of the
drop, $< 10 \%$ in all the simulations reported here. (Considering
different viscosities for drop and vapor leads to the same
conclusion.)  To tie in with this we would expect the extra
dissipation due to the presence of the drop to be relatively
small. The magnitude of $x$-component of the viscous term in different
regions of the drop is shown in \fig{fig:vtx}(b). The dissipation is
increased near the interface and, particularly, near the contact
line. However this increase represents only a few percent of the total
viscous dissipation; the rest being due to the Poiseuille flow in the
bulk.

Thus the drop is effectively convected along by the flow %Poiseuille
field and its velocity is determined by its position within this
flow. If the drop radius or the contact angle are larger, the drop
centre of mass lies at a higher $z$ position within the flow and the
drop therefore moves faster.

\begin{figure}
\begin{center}
\begin{tabular}{ccc}
 \epsfig{file=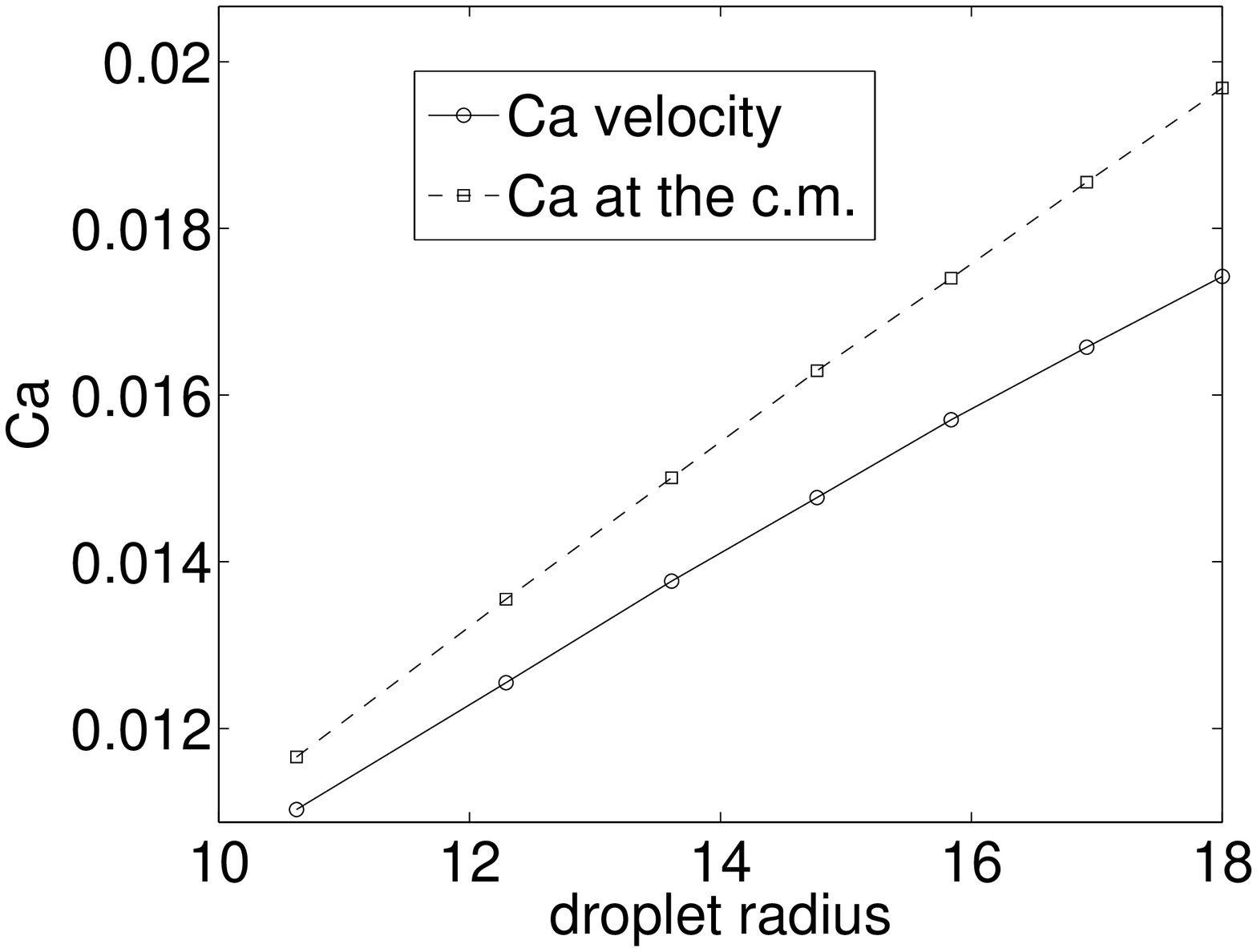,width=4.3cm} & 
 \epsfig{file=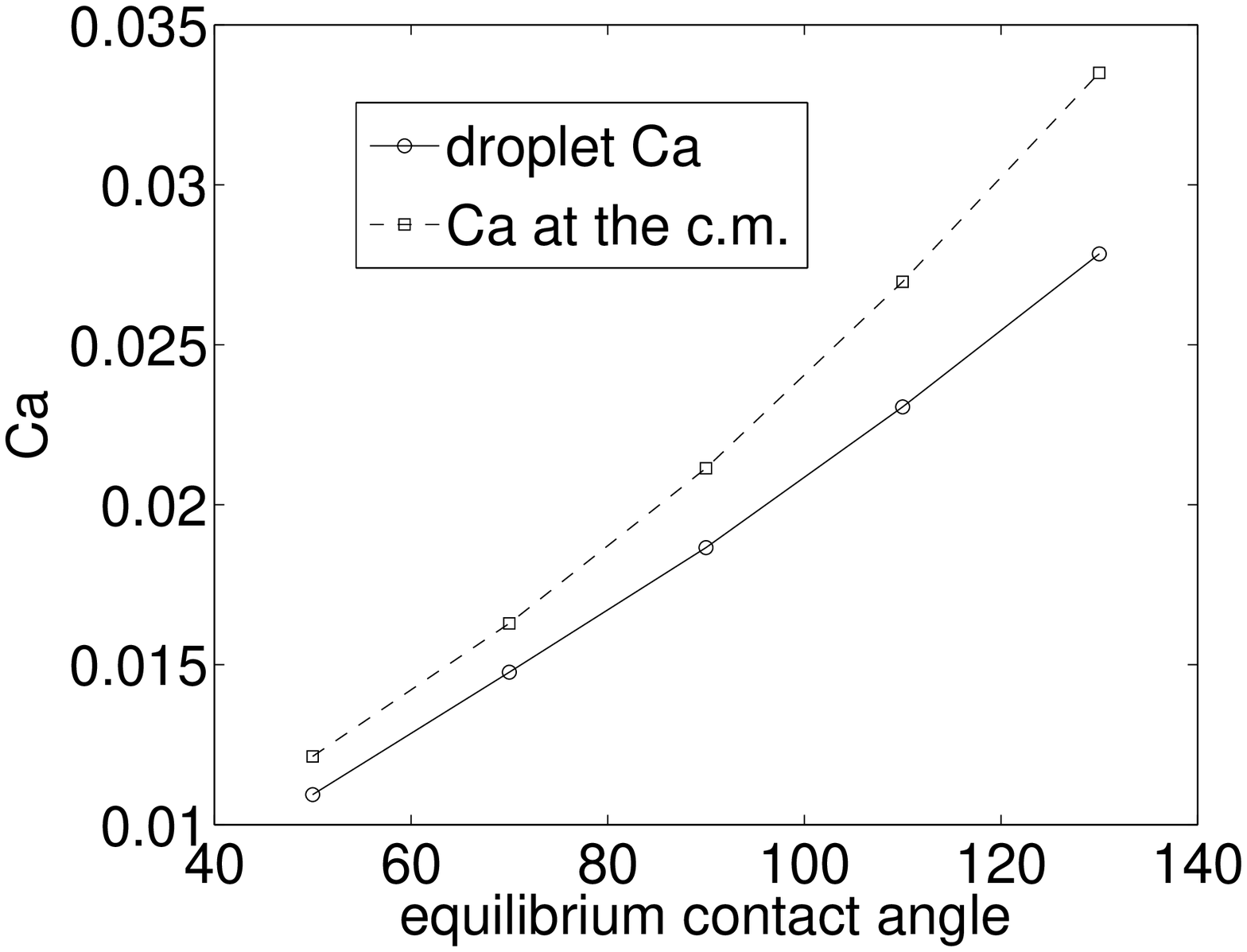,width=4.3cm} & 
 \epsfig{file=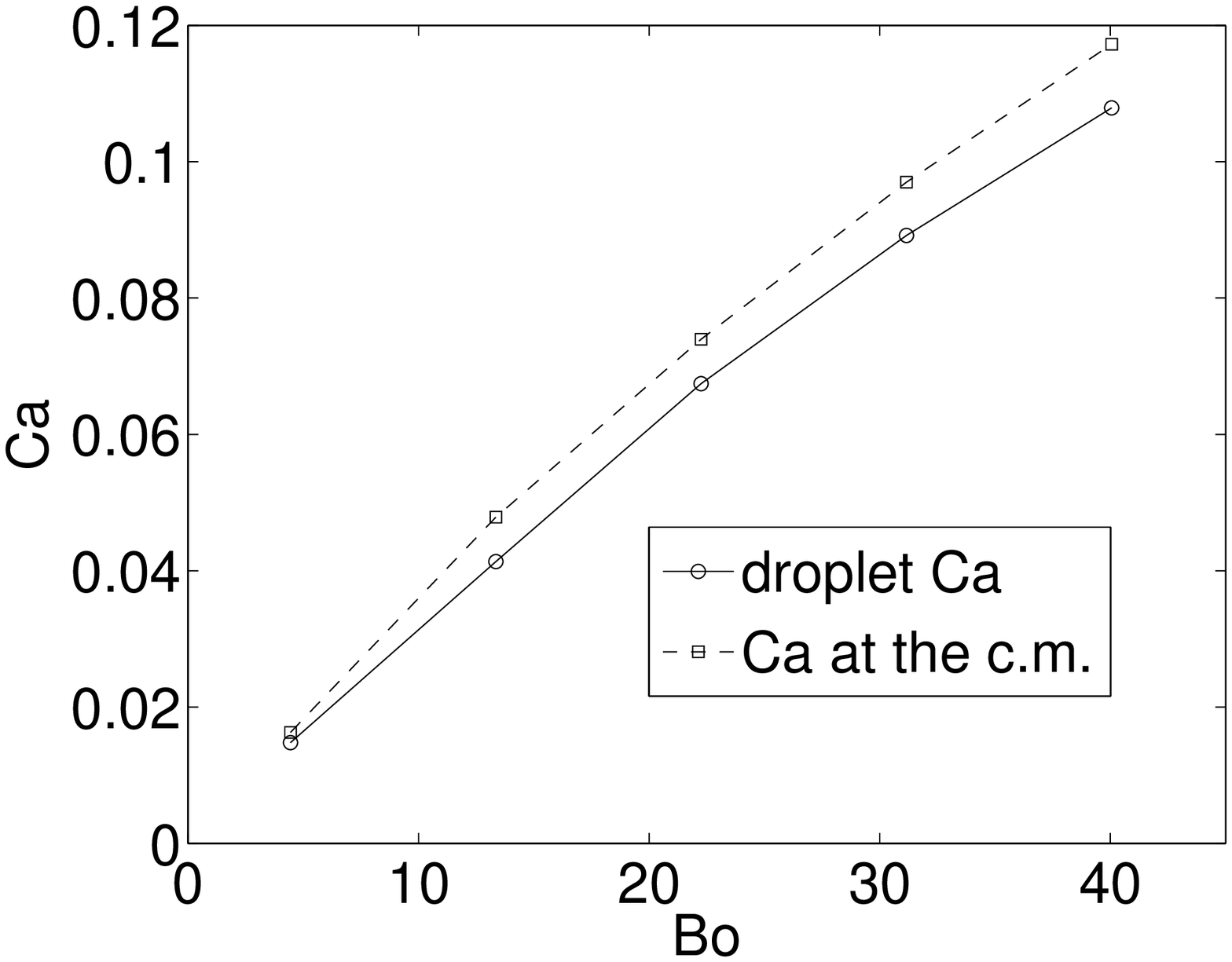,width=4.2cm} \\
(a) & (b) & (c)
\end{tabular}
\end{center}
\caption{Variation of drop capillary number with (a) radius $R$ (in
  simulation units) for $Bo=6.41$ and $\theta_0=70^o$; (b) contact
  angle for $R=15$ and $Bo=6.41$; (c) driving force for $R=15$ and
  $\theta_0=70^o$.}
\label{fig:figs1}
\end{figure}

This is illustrated in \fig{fig:figs1} where we compare the velocity
of the drop and the speed of the flow at the position of its centre of
mass for different drop radii, equilibrium contact angles, and imposed
forces. There is close qualitative and reasonable quantitative ($\sim
10 \%$) agreement. An exact match is not to be
expected because of the enhanced dissipation around the interface.

%\subsection{Motion of a suspended drop on a superhydrophobic surface}

We are now in a position to explain the results in
\fig{fig:velOnPosts}(a), in particular that the velocity of the drop
depends on $w/d$, the ratio of the post width $w$ and the post spacing
$d$. This follows because the velocity of the drop is primarily
determined by its position within the flow field which depends on the
apparent contact angle. This, in turn, is determined, through Cassie's
law by $w/d$, for a given value of the flat surface contact angle.

This can be made more quantitative if we approximate the shape of the
suspended drop as a spherical cap. Cassie's law relating the flat
surface contact angle $\theta_0$ to the observed contact angle on the
superhydrophobic substrate $\theta$ is
\begin{equation}
\cos \theta = (w/d)^2 (\cos \theta_0 +1)-1.
\end{equation}
The position of the centre of mass of the drop is
\begin{equation}
\bar{z}=\bar{R}\left\{\frac{3-8\cos \theta+6\cos^2 \theta-\cos^4 \theta}
{8-12 \cos \theta +4\cos^3 \theta}\right\} \quad \textrm{with} \quad
\bar{R}=R\left\{\frac{4}{2-3 \cos \theta +\cos^3 \theta}\right\}^{1/3}
\end{equation}
where $\bar{R}$ is the radius of the spherical cap.
%where $\bar{R}$, the radius of the spherical cap, is related to the
%radius $R$ of the equivalent spherical drop by
%
%\begin{equation}
%\bar{R}=\left\{\frac{4}{2-3 \cos \theta +\cos^3 \theta}\right\}^{1/3}R
%\end{equation} 
%
For Poiseuille flow the dimensionless velocity  is
%
%\begin{equation}
%v_x=\frac{F_x D}{2 \nu} \left( \bar{z} - \frac{\bar{z}^2}{D} \right)
%\label{eq:udrop}
%\end{equation}
\begin{equation}
v_x^*=\frac{4Bo}{(4/3 \pi)^{2/3}} \left( \bar{z} - \frac{\bar{z}^2}{D_c} \right)
\label{eq:udrop}
\end{equation}
where $D_c$ is the diameter of the channel and $\nu=0.1$ is the
kinematic viscosity.

Table 1 compares the numerical results to \eq{eq:udrop} with
$D_c=L_z-h$. As before the formula agrees qualitatively with the
simulations and predicts a velocity $\sim 10\%$ higher than that
measured. For smaller values of $w/d$ the numerical and analytic
results are closer, but this is because the effective width of the
channel decreases as the number of posts increase.

\begin{table}
\begin{center}
\begin{tabular}{c|c|c}
$w/d$  &  simulated $Ca$ (from \fig{fig:velOnPosts})
  &  estimated velocity, $v_x^*$ (from \eq{eq:udrop})\\
%&(simulation units)&(simulation units)\\
\hline
%1     &  $3.48 \cdot 10^{-4}$ & $4.07 \cdot 10^{-4}$  \\ 
1     &  $0.0076$ & $0.0089$  \\ 
%0.73   &  $4.10 \cdot 10^{-4}$ & $4.68 \cdot 10^{-4}$  \\
0.73   &  $0.0089$ & $0.0102$  \\
%0.6   &  $4.43 \cdot 10^{-4}$ & $4.92 \cdot 10^{-4}$   \\
0.6   &  $0.0096$ & $0.0107$   \\
%0.4   &  $4.78 \cdot 10^{-4}$ & $5.20 \cdot 10^{-4}$  \\
0.4   &  $0.0104$ & $0.0113$  \\
0.25   &  $0.0114$ & $0.0116$  \\ \hline
\end{tabular}
\end{center}
\caption{Comparison of simulated and estimated velocity for a drop
moving across a superhydrophobic surface.}
\end{table}

%\begin{figure}
%\begin{center}
%\epsfig{file=vel2.eps,width=5.5cm}
%\epsfig{file=fitSusp.eps,width=5.5cm}
%\end{center}
%\caption{(Fit (solid line) of the the results (circles) from
%  \fig{fig:velOnPosts}(a) with \eq{eq:udrop} considering $D=L_z$ ($w/d
%  < 0.4$) and $D=L_z-h$ ($w/d \geq 0.4$). The dashed line is a linear
%  fit of the simulations. The center of mass velocity of a sphere
%  (square) located at $z=35$, moving on a cushion of air.}
%\label{fig:velEstimated}
%\end{figure}

%\subsection{Motion around a single post}

\begin{figure}
\begin{center}
\begin{tabular}{ccc}
\epsfig{file=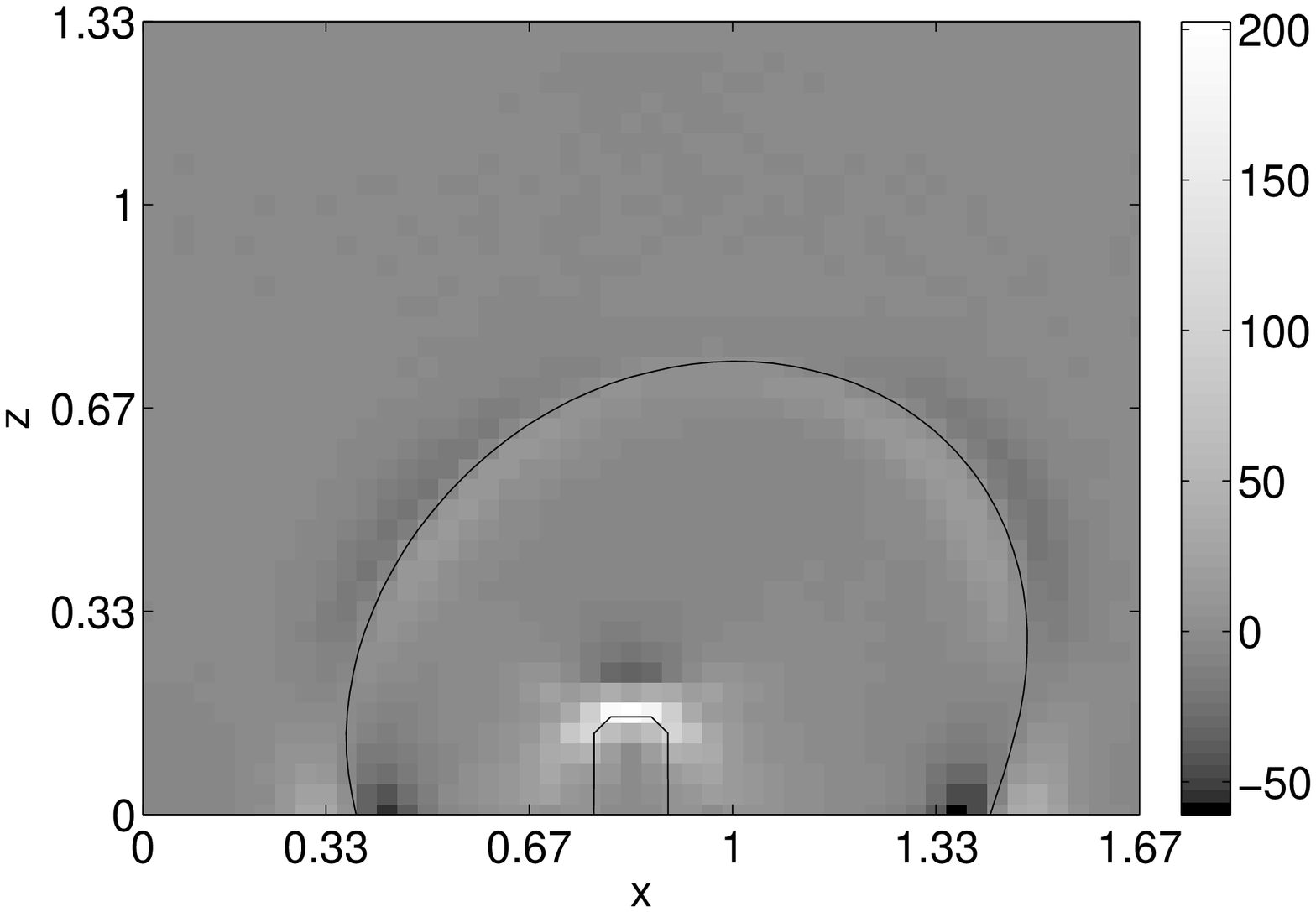,width=6.3cm} & &
\epsfig{file=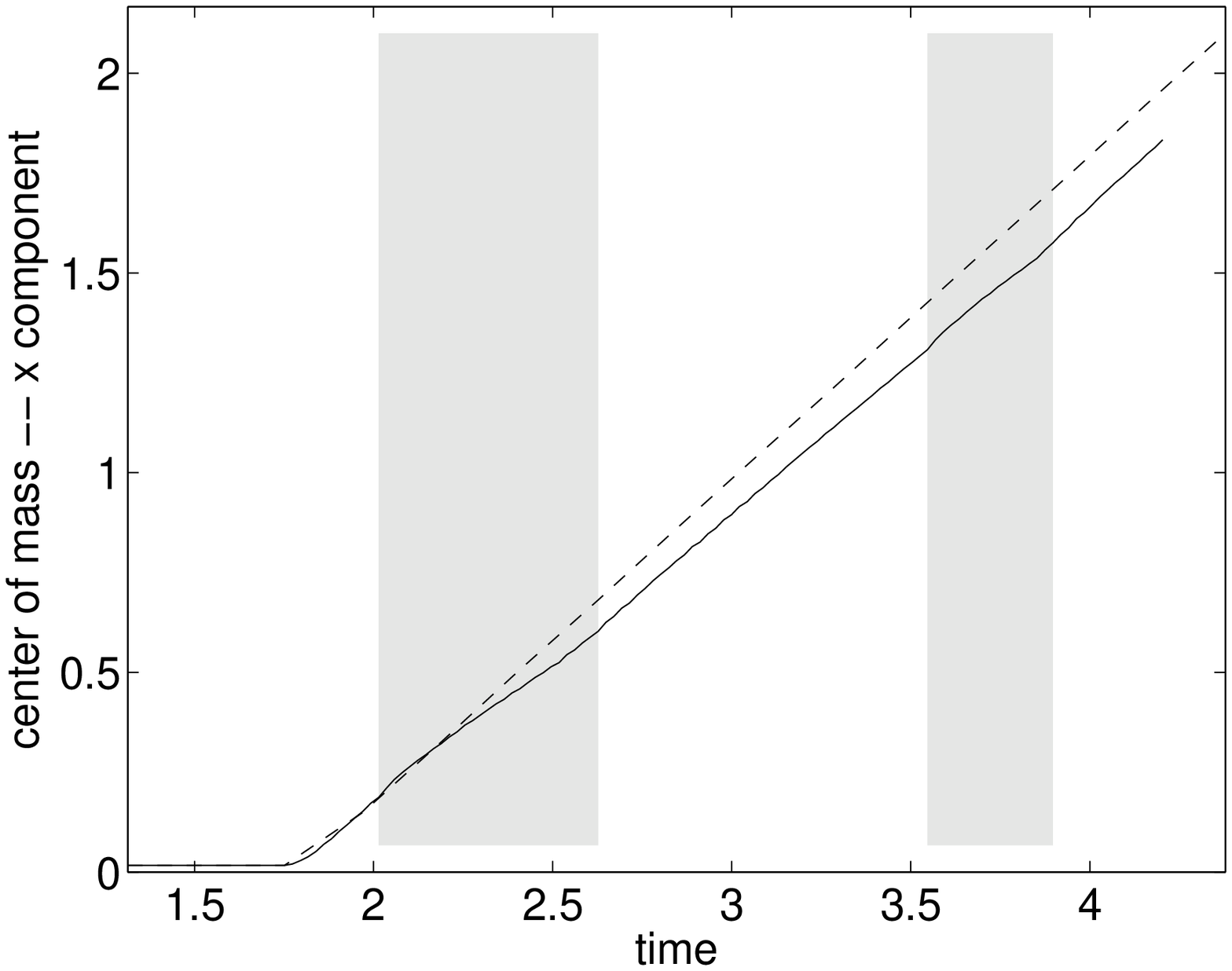,width=5.4cm} \\
(a) && (b)
\end{tabular} 
\end{center}
\caption{(a) $x$-component of the viscous term in the
  Navier-Stokes equation (in dimensionless unit, scaled by
  $V^{(2/3)}/\gamma$) for a drop moving across a surface decorated
  with a single post. Cross section at $y=L_y/2$. (b) Comparison of
  the variation of the position of the centre of mass with time of two
  drops. Dashed line: drop moving across a flat surface. Full line:
  drop moving across a surface decorated by a post. Both length and
  time are dimensionless scaled by the drop diameter $D$ and $\nu n
  D/\gamma$ respectively. The grey areas between $t=2$ and $t=2.6$ and between
   $t=3.5$ and $t=3.9$ correspond to the advancing and to the receding 
contact lines
  being on the post respectively. $\theta_0=110^o$ on every surface
  site.}
\label{fig:post}
\end{figure}

\Fig{fig:velOnPosts}(b) shows that a different mechanism is in play
for collapsed drops. To understand this we consider flow over a $4
\times 4$ square post of height $5$ centered within a domain of size
$L_x \times L_y \times L_z = 50 \times 40 \times 40$ lattice sites.
The contact angles on the post and on the surface are identical,
$\theta_0=110^o$. No-slip velocity boundary condition are imposed on
every substrate site. A droplet of radius $R=15$ is equilibrated away
from the post. The drop is convected by a Poiseuille flow with
$Bo=4.5$ until it eventually meets the post. Dissipation due to
friction occurs at the top of the post as it distorts the flow field
as shown in \fig{fig:post}(a).
 
\Fig{fig:post}(b) compares the velocity of two drops, one (F) moving
on a flat surface and the other (P) over a post. The velocities cease
to be identical as soon as the drop meets the post. Because of the
distortion of the flow field the velocity of P is lower than that of
F. The velocities become identical again as soon as the drop leaves
the post. The velocity difference decreases as the height of the posts
is decreased.

%\subsection{Motion of collapsed drops on a superhydrophobic surface}

With this information we return to \fig{fig:velOnPosts}(b).
Increasing $w/d$ from $0$ corresponds to inserting posts into the drop
path. As shown above this leads to increased dissipation as the posts
distort the flow field. Hence the velocity drops, the decrease being
greater for the higher posts.

For $w/d$ above $\sim 0.7$, the velocity jumps to a higher value and
then decreases as more posts are inserted. Here the velocities are
identical to those measured for the suspended drop (see
\fig{fig:velOnPosts}(a)). This is because the force drives a
transition from the collapsed to the suspended state when posts cover
a large fraction of the surface. The complicated behaviour for $w/d
\sim 0.5$ occurs because parts of the drop are collapsed and parts
suspended.

% ------------------------------------------------------------
\section{Conclusion}

We have solved the hydrodynamic equations of motion for drops sliding
over superhydrophobic surfaces. For drops suspended on the surfaces
there is an increase in velocity of about $50\%$ as the number of
posts is decreased to zero. We show that the main contribution to this
effect is from the position of the drop in the Poiseuille flow
field. As the number of posts is decreased the contact angle increases
and hence the drop lies, on average, further from the surface. Thus it
is subjected to a higher velocity field and moves more quickly.

For collapsed drops the situation is very different. Here, as posts
are introduced, they impede the drop and its velocity falls. For a
large number of posts, as the drop is pushed, it prefers to revert to
the suspended state. 

These results are consistent with the few experiments in the
literature~\cite{gogte:05,ou:04}. However the model we solve does
include approximations to make it numerically
tractable and it is important to be aware of these.
In particular interfaces tend to move too easily, because, 
as with all mesoscale two-phase fluid simulations,
simulated interface widths are too wide. The effect of this is that,
given a set of input physical parameters, the simulation drop will
move more quickly than a real drop.  This can be accounted for by a
rescaling of time by, for example, matching the capillary number at
which the moving drop develops a corner in its trailing edge to
experiment~\cite{podgorski:01}. Moreover it is necessary to choose
boundary condition for the model. We have taken a non-slip condition
on the velocity on every surface site: although there is evidence for
increased local slip on hydrophobic surfaces this is still on the nano
metre level and so would not show up on the scale of these
simulations.

The drops we consider are in the sliding, rather than the rolling
regime. Most experiments to date have been on highly viscous drops
which roll~\cite{richard:99}, but it is likely that there will soon be
experiments on drops with lower viscosity which will allow interesting
comparisons to the simulations reported here.

% ============================================================
\section{Acknowledgment}

AD acknowledges the support of the EC IMAGE-IN project
GR1D-CT-2002-00663. We thank H.~Kusumaatmaja for useful comments on
the draft.

%\def\biblioP{/Users/dupuisa/Tex/Bib/} 
%\bibliographystyle{unsrt}
%\bibliography{\biblioP dupuis,\biblioP lattice-gas,\biblioP wetting}

%\end{document}

\end{document}